\documentclass[12pt]{article}
\usepackage{amsmath}
\usepackage{graphicx}
\usepackage{natbib}
\usepackage{url} % not crucial - just used below for the URL 

\usepackage{times}
\usepackage{bm}
\usepackage{natbib}
\usepackage{amssymb}
\usepackage{caption}
\usepackage[plain,noend]{algorithm2e}
\usepackage{xcolor}

\newtheorem{assumption}{Assumption}[section]
\newtheorem{remark}{Remark}[section]
\newtheorem{theorem}{Theorem}[section]
\newtheorem{proposition}{Proposition}[section]

\newcommand{\blind}{0}

\begin{document}

\def\spacingset#1{\renewcommand{\baselinestretch}%
{#1}\small\normalsize} \spacingset{1}

%%%%%%%%%%%%%%%%%%%%%%%%%%%%%%%%%%%%%%%%%%%%%%%%%%%%%%%%%%%%%%%%%%%%%%%%%%%%%%

\if0\blind
{
  \title{\bf Kendall's Tau for Functional Data Analysis}
  \author{Sneha Jadhav\hspace{.2cm}\\	
    Department of Mathematics \& Statistics, Wake Forest University\\
    and \\
    Shuangge Ma \\
    Department of Biostatistics, Yale University}
\date{}
  \maketitle
} \fi

\if1\blind
{
  \bigskip
  \bigskip
  \bigskip
  \begin{center}
    {\LARGE\bf Title}
\end{center}
  \medskip
} \fi

\bigskip
\begin{abstract}
We treat the problem of testing for association between a functional variable belonging to a Hilbert space and a scalar variable. 	Particularly, we propose a distribution-free test statistic based on Kendall's Tau which is one of the most popular methods to determine the association between two random variables. The distribution of the test statistic under the null hypothesis of independence is established using the theory of U-statistics taking values in a Hilbert space. We also consider the case when the functional data is sparsely observed, a situation that arises in many applications. Simulations show that the proposed method outperforms the alternatives under several conditions demonstrating the robustness of our approach. We provide data applications that further consolidate the utility of our method.  

\end{abstract}

\noindent%
\vfill

\newpage
\spacingset{1.5} % DON'T change the spacing!
\begin{center}{\section{INTRODUCTION}}
\end{center}
\label{sec:intro}

A fundamental problem that arises often in statistics is to determine whether a pair of random variables are independent or are related to each other. This problem gets increasingly difficult as the technological advancement leads to variables of a complex nature. An important example of such a complex variable is a random function. In this article, we focus on testing for association between a scalar variable and a random function.  In many scientific fields, functional data routinely arise, and accordingly, extensive statistical developments have been carried out. In the literature, the most common way for the association test is via linear modeling, where the response variable is continuously distributed and the regressor is functional (of note, there are also studies that reverse the roles). Examples include \cite{cardot2003testing,horvath2012inference}, which develop testing for functional linear models based on the cross covariance operator. \cite{muller2005, jadhav2017dependent} develop a testing framework for generalized functional linear models. \cite{kong2016classical} extends the Wald, score, likelihood ratio, and F tests to functional linear models. \cite{swihart2014restricted} develops a restricted likelihood ratio based test for functional linear models. \cite{reimherr2014functional} conducts a functional-on-scalar regression for testing associations in genetic studies. \cite{goldsmith2011penalized} use mixed models for constructing confidence intervals in penalized functional regression. \cite{lei2014adaptive,su2017hypothesis} develop a method of ordering and selecting principal components to construct tests in functional linear models. Our literature review suggests that most of the existing methods are parametric (based on functional linear models) and limited to checking for linear association. There are also a handful of nonparametric regression based approaches. For example \cite{cardot2007no}. Only a few methods allow for sparsely observed and noisy functional data.  As will be evident from the simulations, the performance of these methods is quite sensitive to their model assumptions. We aim to overcome these shortcomings by building a more robust nonparametric rank based method.

In this article, we take a significantly different strategy of addressing the problem of association detection directly, instead of first modeling the relation between the two variables and then determining association. Our strategy is based on U-statistics \citep{hoeffding1992class} which 
represent an important class of statistics having wide applications in the areas of estimation and inference.  A popular application in the context of association testing is Kendall's Tau which is based on the concept of concordance. Briefly, suppose $(X_1,Y_1),\,(X_2,Y_2) $ are identical and independent copies of random vector $(X,Y)$. Variables $X$ and $Y$ are in concordance if $X_1<X_2$ and $Y_1<Y_2$ or  $X_1>X_2$ and $Y_1>Y_2$. Equivalently, concordance is determined by $\tau= I[(X_1-X_2)(Y_1-Y_2)>0]$, which is the crux of the Kendall's Tau based test. We extend this idea to the functional framework to propose a U-statistic that takes values in Hilbert space or UH-statistic. The central limit theorem of UH-statistics is used to determine the distribution of the test statistic under the null hypothesis. We also address the case when functional data is sparse by using the conditioning step developed in  \cite{yao2005functional}. Akin to Kendall's Tau, our method has power against the alternative of monotone association. Even so, as linear correlation is also a case of monotone association, our non-parametric method has wider applicability than most existing alternatives.

Simulations demonstrate the advantages of our method over alternatives based on functional linear regression under varying conditions. Our method is better than the parametric alternatives at detecting linear relations when the dimension (number of components in a basis expansion) of the functional data is increased. In addition, the proposed approach is also advantageous with its simplicity -- it directly addresses association testing and avoids complex estimations. All of these factors give our method an edge for real applications. We consider two such applications. In the first application, the functional variable is the cerebral white matter tract of a person with multiple sclerosis (commonly known as MS) and the scalar variable is their cognitive ability. The second application consists of price trajectory of Palm M515 personal digital assistants (PDA) and their opening price in a 7-day auction on e-Bay. In both the applications, there is no reason to believe that the relation is linear and thus, may not be captured by methods based on linear models and hence, it is more reasonable to use the proposed approach.

%There are several applications in which we wish to determine the association between functional and a scalar variable. We consider two such applications. In the first application, the functional variable is the cerebral white matter tract of a person with multiple sclerosis and the scalar variable is their cognitive ability. The second application consists of price trajectory of Palm M515 personal digital assistants (PDA) and their opening price in a 7-day auction on e-Bay. In both the applications, there is no reason to believe that the relation is linear and thus, may not be captured by methods based on linear models. Our proposed approach may be more useful in these cases. Simulations demonstrate the advantages of our method over alternatives based on functional linear regression under varying conditions. An important advantage of our method is that it is better than the parametric alternatives at detecting linear relations when the dimension (number of components in a basis expansion) of the functional data is increased. In addition, the proposed approach  is also advantageous with its simplicity -- it directly addresses association testing and avoids complex estimations.  

The rest of the article is organized as follows. In section 2, we introduce the test statistic, establish its properties and discuss its implementation. We also cover the case of sparse functional data. Section 3 consists of simulation studies that compare the performance of the proposed method with alternatives based on the functional linear model. In section 4, applications to diffusion tensor imaging and Ebay auction data are presented. The proofs are deferred to the Appendix.

\begin{center}{\section{METHODOLOGY}}
	\end{center}
\subsection{Functional Data}
The data consists realizations of random variables $(Y_i, X_i),\,i=1,...,n$ which are independent and have distribution identical to $(Y,X)$, where  $X$ is a smooth curve $\in L^2(I)$ and $ Y \in \mathbb{R}$ has a continuous distribution. We consider the interval $I=[0,1]$ without any loss of generality. This Hilbert space is equipped with the norm $ \|X\|^2=\int_I X^2(t)dt$ and inner product $\langle X_1 , X_2\rangle =\int_I X_1(t)X_2(t)dt$. Note that all integrals hence onwards are taken over $I$ unless stated otherwise. Let $A$ be a subset of $I$ with positive Lebesgue measure. 
Our interest is to determine whether there is any association between $X$ and $Y$.  Formally, we wish to test the following hypothesis:
\begin{align} \label{hypo}
&\{H_0:  Y \text{ and } X(t) \text{ have no association (independent), } \forall t \in I\} \nonumber \\
&\quad \quad \quad \quad \text{ vs.} \\
& \{H_1:  Y \text{ and }  X(t) \text{  have a monotone association},\, \forall t \in A \} \nonumber
\end{align}
Thus, association is detected on a set of positive measure. To test this hypothesis consider the following  UH-statistic
\begin{equation}\label{u1}
U_n(t)= {n \choose 2}^{-1}  \sum_{1\leq i<j\leq n} \mathcal{I}[(Y_i-Y_j) (X_i(t)-X_j(t))>0]-0.5,
\end{equation}
where $\mathcal{I}(\cdot)$ is an indicator function. As, $E_{H_0}(\mathcal{I}[(Y_i-Y_j) (X_i(t)-X_j(t))>0])=0.5,$ the UH-statistic is centered under the null. We propose the test statistic
\begin{equation}\label{test}
T= \| U_n\|^2 .
\end{equation}
For asymptotic results, we make the following assumption:

\begin{assumption}
	$E\|X\|^4 < \infty. $
\end{assumption}

This assumption is commonly found in the functional data literature which guarantees the consistency of the sample covariance operator. In the rest of the work, all the expectations are taken under the the null hypothesis ($H_0$),  $G.P.(0,C)$ denotes a centered Gaussian random function with covariance operator $C$, $Z=(X,Y)$, small letters indicate fixed values for random variables and all limits are as $n \to \infty$.
\begin{theorem}\label{thm1}
	Under $H_0$, $T \to \int \Gamma^2(t) dt,	$ where $\Gamma \sim G.P.(0,\mathcal{C})$ and $\mathcal{C}$ is the covariance operator of the random function $V(t)=E[ \mathcal{I}\{(Y_1-Y) (X_1(t)-X(t))>0\}\mid Z_1]-0.5$. Recall that $Z_1=(Y_1,X_1)$ is an independent copy of the random variable $Z=(Y,X)$ .	
\end{theorem}
This theorem is a consequence of Theorem 4.2.2 from \cite{borovskikh1996u}, that establishes the central limit theorem for U-statistics in Hilbert spaces. Karhunen-Lo\`{e}ve expansion yields $\int \Gamma^2(t)dt=\sum_{k=1}^{\infty}\lambda_kN_k^2, $ where, $N_k$ are independent standard normal variables, $\lambda_1 \geq \lambda_2\geq... $ are eigenvalues of the covariance operator $\mathcal{C}$. Thus, an approximation of the asymptotic distribution of $T$ under $H_0$ can be obtained from the estimate of the eigenvalues of the operator $\mathcal{C}$. This approximation is then used to obtain  $ z_{\alpha}, $ which is the $100(1-\alpha)^{th}$ percentile of the distribution of $\int \Gamma^2(t) dt$. Thus, a test that rejects $H_0$ if the test statistic $T> z_{\alpha}$ is of asymptotic size $\alpha$.

 %For clarity, we would like to point out that the covariance operator of a random function $U $ is given by $\mathcal{M}(u_0)= E\{\langle U,u_0 \rangle U\},\, u_0 \in L^2 $. For a sample $U_1,...,U_n$ the empirical estimate of this operator is given as $\widehat{\mathcal{M}}(u_0)=\sum_{i=1}^{n} \langle U_i,u_0\rangle U_i$.

We now focus on obtaining the estimates of the eigenvalues $\lambda_k,\,k\geq 1$, or equivalently estimating the operator $\mathcal{C}$.  Let $V_i(t)=E[ \mathcal{I}\{(y_i-Y) (x_i(t)-X(t))>0\}\mid Z_i=z_i]-0.5=E[ \mathcal{I}\{(y_i-Y) (x_i(t)-X(t))>0\}]-0.5,\,i=1,...,n $. Note that the functions $V_i$ are independent copies of the function $V$. Given that $E(V_i)=0$, the estimate of the covariance operator $\mathcal{C}$   based on $V_i's$ is ${\mathcal{C}}_v(x)=n^{-1}\sum_{i=1}^{n} \langle V_i , x \rangle V_i. $ However, the variables $V_i$ are not directly observed and must be estimated. The empirical estimate of $V_i$ is $\widehat{V}_i(t)=W_i(t)=n^{-1}\sum_{j=1}^{n} I[(y_i-Y_j)(x_i(t)-X_j(t))>0]-0.5.$ Denote $\mathcal{C}_w(x)= n^{-1}\sum_{i=1}^{n} \langle W_i , x \rangle W_i$. As the $W_i's$ are directly determined from the observed data, the asymptotic distribution of the test statistic $T$ can be determined  if the operator $\mathcal{C}_w$ provides a reasonable approximation for $\mathcal{C}$. To investigate this, we make use of the Hilbert-Schmidt norm for operators given by $ \|\mathcal{C}\|_S^2=\sum_{k=1}^{\infty} \|\mathcal{C}(e_k)\|^2,$ where $e_k,\,k\geq1$ form an orthonormal basis in $L^2[0,1]$. 

\begin{theorem}\label{thm2}
	$E\| \mathcal{C}_w- \mathcal{C}\|^2_{S} \to 0$	
	
\end{theorem}
The proof is given in the Appendix. Using this result, the eigenvalues of  $\mathcal{C}_w$ are the estimates of the eigenvalues of $\mathcal{C}$. The sum $\sum_{k=1}^{d}\widehat{\lambda}_kN_k^2 $ is used to approximate $\sum_{k=1}^{\infty}\lambda_kN_k^2, $ where $d$ is large enough. We choose $d$ to be the number of eigenvalues that explain 95\% variance. %\textcolor{red}{The details of selection of $d$ are given in section 3.}

\begin{remark}
	It is not practically feasible to observe the entire curve, but we can observe its realizations on a dense grid of points. Several smoothing techniques are available to recover the underlying curve \citep{ramsay2006functional, zhu2014structured}.  We take the commonly used  approach to treat these recovered curves as the observed data.
\end{remark}

\subsection {Sparsely observed Functional Data}
%In this section we propose two approaches,  Method I is based on the PACE technique developed by \cite{yao2005functional} and Method II is based on the work of \cite{james2000principal}. Though both of these works are based on sparse functional data, their framework is quite different. Accordingly, the model and assumptions depend on the approach under consideration. 

%\subsection{Method I}
We use the framework from \cite{yao2005functional} to extend our method to the case of sparse data. We provide information necessary to implement the method and refer the reader to the aforementioned work for greater details of the underlying assumptions. The functional data is observed over a sparse grid of irregular points. This sparsely observed data is modeled as noisy realizations of independent smooth random functions $X_i,\,i=1,...,n$ with the same distribution as a random function $X$ . Thus, the data consists of  observations $G_{ij}$, where $G_{ij}=X_i(T_{ij})+ \varepsilon_{ij},\,T_{ij} \in I,$ $j=1,...,N_i$ and $i=1,...,n,$ where $T_{ij} $ and $N_i $ are each identical and independent random variables, $E(\varepsilon_{ij})=0 $ and $var(\varepsilon_{ij})=\sigma^2$. %We refer the reader to \cite{yao2005functional} for further details on the assumptions regarding the involved variables. We would like to point out that variables $N_i$ and $T_{ij}$ reflect the irregular and sparse nature of the design points.
Though the observations for each curve are sparse, we assume that the design points $T_{ij}$ are sufficiently dense when pooled together. Denote the mean function $E(X(t))$ by $\mu(t)$, the eigenvalues and corresponding eigenfunctions of the covariance operator of $X(\cdot)$ by $\gamma_k$ and $ \phi_k(t),\, k\geq 1$. The principal component scores are $\xi_{ik}=\int \phi_k(t)(X_i(t)-\mu(t))dt$. We assume that the score $\xi_{ik}$ and error $\varepsilon_{ij}$ are jointly Gaussian. Let $\widetilde{\mathbf{X}}_i=(X_i(T_{i1}),...,X_i(T_{iN_i}))^{T},\,\widetilde{\mathbf{G}}_i=(G_{i1},...,G_{iN_i})^{T},\,\boldsymbol{\mu}_i=(\mu_i(T_{i1}),...,\mu_i(T_{iN_i}))^{T}$ and $\boldsymbol{\phi}_{ik}= (\phi_k(T_{i1}),...,\phi_k(T_{iN_i}))^{T}.$ The best predictor of $\xi_{ik}$ is the conditional expectation $$\widetilde{\xi}_{ik}=E(\xi_{ik} \mid \widetilde{\mathbf{G}}_i,T_{ij})=\gamma_k\boldsymbol{\phi}_{ik}^T\boldsymbol{\Sigma}_{\mathbf{G}_i}^{-1}(\widetilde{\mathbf{G}}_i-\boldsymbol{\mu}_i), $$
where, the $ (j,l)$ entry of  matrix  $ \boldsymbol{\Sigma}_{\mathbf{G}_i} $ is $cov(X(T_{ij}),X(T_{il}))+\sigma^2\delta_{jl},\,\delta_{jl}=1$ if $l=j$ and $\delta_{jl}=0$ otherwise. We use the predicted scores $ \widetilde{\xi}_{ik} $ to construct identical and independent functions $\widetilde{X}_i(t)=\mu(t)+\sum_{k=1}^{\infty}\widetilde{\xi}_{ik}\phi_k(t)$ which are used to construct the statistic needed to test hypothesis  \eqref{hypo}. 

Let
\begin{equation*}
\widetilde{U}_n(t)= {n \choose 2}^{-1} \sum_{i<j}^{n} \mathcal{I}[(Y_i-Y_j) (\widetilde{X}_i(t)-\widetilde{X}_j(t))>0]-0.5,
\end{equation*}
where, $I(\cdot)$ is an indicator function. The statistic is
\begin{equation}\label{test2}
\widetilde{T}= \|\widetilde{U}_n   \| ^2.
\end{equation}
Note that the above test statistic determines whether $Y$ has no association with the function $\widetilde{X}$. In light of Theorem 3 and Theorem 4 from \cite{yao2005functional}, and the simulation results in the next section it is reasonable to use statistic \eqref{test2} for testing hypothesis \eqref{hypo}. To avoid repetition, we state only some of the assumptions required for the results in this subsection. 
% However, given that the design points are dense when pooled together it is reasonable to use \eqref{test2} for testing \eqref{hypo}. 

\begin{assumption}
	Let $N_i$ be realizations of a random variable $N$. Then, $N $ is a positive discrete random variable with $E(N)<\infty  $ and $P(N>1)>0. $  
\end{assumption}

\begin{assumption}
	Let $J_i \subseteq \{ 1,...,N_i\}$. Then, $ \{T_{ij}:j \in J_i\}; \{G_{ij}:j \in J_i \}   $ is independent of $N_i$.
\end{assumption}

\begin{assumption}
	$\xi_{ik}$ and $\varepsilon_{ij}$ are jointly Gaussian. 
\end{assumption}

\begin{theorem}\label{thm3}
	Let $\widetilde{X}$ be a random function that has the same distribution as $\widetilde{X}_i$'s. If $H_0$ holds, $\widetilde{T} \to \int \widetilde{\Gamma}^2(t) dt,	$ where $\widetilde{\Gamma} \sim G.P.(0,\widetilde{\mathcal{C}})$ and $\widetilde{\mathcal{C}}$ is given by the covariance operator of the random function $\widetilde{V}(t)=E[\mathcal{I}\{(Y_1-Y) (\widetilde{X}_1(t)-\widetilde{X}(t))>0\}\mid \widetilde{Z}_1]-0.5,$ where $(Y_1,\widetilde{X}_1)$ is an independent copy of the random variable $(Y,\widetilde{X})$ and $\widetilde{Z}=(Y_i,\widetilde{X}_i) $.	
\end{theorem}
This follows directly from the central limit theorem for U-statistics in Hilbert spaces. To compute the test statistic $\widetilde{T},$ several involved components need to be estimated. In order to obtain estimate of functions $\widetilde{X}_i$, it is necessary to determine estimates of the involved parameters given by $\widehat{\phi}_k,\,\widehat{\xi}_{ij}$ and  $\widehat{\mu}$. Then, the corresponding estimated functions are $\widehat{X}_i(t)=\widehat{\mu}(t)+\sum_{k=1}^{\infty}\widehat{\xi}_{ik}\widehat{\phi}_k(t).$   The projection of function $\widetilde{X}_i$ along the first $K$ eigenfunctions is $\widetilde{X}_i^K(t)=\mu(t)+\sum_{k=1}^{K}\widetilde{\xi}_{ik}\phi_k(t)$ and that of $\widehat{X}_i$ are $\widehat{X}_i^K(t)=\widehat{\mu}(t)+\sum_{k=1}^{K}\widehat{\xi}_{ik}\widehat{\phi}_k(t).$  Cross-validation is used to determine the value for $K$. We refer the reader to \cite{yao2005functional} to see the details regarding these estimates. We would also like to state the Theorem 3 from this paper:
for all $t \in I$, $\widehat{X}_i^K(t) \to \widetilde{X}_i(t),\, n \to\infty,\, K \to \infty $ in probability. Let $\widehat{U}_n^K= {n \choose 2}^{-1} \sum_{i<j}^{n} \mathcal{I}[(Y_i-Y_j) (\widehat{X}_i^K(t)-\widehat{X}_j^K(t))>0]$ and $\widehat{T}^K =\|\widehat{U}_n^K \| ^2.$

\begin{theorem}\label{thm4}
	
	$ \underset{K \to \infty}{lim} \underset{n \to \infty}{lim}  \widehat{U}_n^K(t)-\widetilde{U}_n(t)=0  $ in probability. This implies 
	$ \underset{K \to \infty}{lim} \underset{n \to \infty}{lim}  \widehat{T}_n^K-\widetilde{T}=0$  in probability.
\end{theorem}
The proof of this theorem is given in the Appendix. From Theorems \ref{thm3} and \ref{thm4}, $\widehat{T}^K  \to \int \widetilde{\Gamma}^2(t) dt$. As before, we use the Karhunen-Lo\`{e}ve expansion to obtain  $\int \widetilde{\Gamma}^2(t)dt=\sum_{k=1}^{\infty}\widetilde{\lambda}_kN_k^2, $ where $N_k$ are independent standard normal variables, $\widetilde{\lambda}_1 \geq \widetilde{\lambda}_2\geq... $ are eigenvalues of the covariance operator $\widetilde{\mathcal{C}}$.  The estimation of the covariance operator $\widetilde{\mathcal{C}} $ and its eigenvalues is similar to the previous dense case. Let $W_i^K(t)=n^{-1}\sum_{j=1}^{n} I[(y_i-Y_j)(\widehat{x}_i^K(t)-\widehat{X}^K_j(t))>0]-0.5.$ Denote $\widetilde{\mathcal{C}}^K(x)= n^{-1}\sum_{i=1}^{n} \langle W_i^K , x \rangle W_i^K$. The asymptotic distribution of $\widehat{T}^K$ can be determined using the eigenvalues of $\widetilde{\mathcal{C}}^K$ if $\|\widetilde{\mathcal{C}}^K-\widetilde{\mathcal{C}}\|_S$ is small.

\begin{theorem}\label{thm5}
	$E\| \widetilde{\mathcal{C}}^K- \widetilde{\mathcal{C}}\|^2_{S} \to 0$ as $ K \to \infty,\, n \to \infty$	
	
\end{theorem}
The proof of this theorem is similar to that of Theorem \ref{thm2} and Theorem \ref{thm4}. Though we make Gaussian assumption, simulations show that the test is fairly robust to this violation.

\begin{center}
	{\section{SIMULATION}}
\end{center}	

In Simulation I, we consider the scenario where the functional data is observed on a regular and a relatively dense grid. To investigate our method throughly, we consider different models for generating the scalar $Y$. A sample of $n$ functions is generated using basis functions $\{ \rho_j,\,j\geq 1\}$ and model $X(t)=\sum_{k=1}^{p}\varepsilon_{k} \rho_{k}(t)$, where $ \varepsilon_j$'s are independent and have an exponential distribution with rate $2$, and observed on $t_1,...,t_{20}$, which are equally spaced on $[0, 1]$. The effect function is $\beta(t)= \sum_{k=1}^{p} \beta_k\rho_{k}(t),$ where $\beta_k=k/2, k\ge 1$. Denote a standard normal variable by $N(0,1)$ and an exponential variable with rate $a$ by $exp(a) $.  To generate $Y$, we consider the following cases:\\
\underline{Case 1}: Fourier basis is used to generate $X$ and  $Y = \delta\int X(t)\beta(t)dt+ N(0,1).$ This is a typical set-up for functional linear regression, where the relationship is linear. \\
\underline{Case 2}:  Fourier basis is used to generate $X$ and $Y=	\delta\int X(t)\beta(t)dt+exp(2).$ In this case, the relationship is linear but the error follows atypical exponential distribution. \\
%\underline{Case 3} We generated processes $U$ in the same way as $X$ given above and additionally, center $U$ so that it's mean function is $0$. Let  $X_1(t)=X(t)+U(t)$. Let $e=N(0,1)-\int U(t)\beta(t)dt$, then $Y=\int X_1(t)\beta(t)dt+e.$ Note that in this case the regressor variable $X_1$ and the error $e$ are correlated violating an important assumption of functional linear regression models that they are uncorrelated. \\
\underline{Case 3}:  For this case, we use a monomial basis. $Y=\int (0.001)^{X(t)}dt + N(0,0.1).$ The relationship between $Y$ and regressor variable $X$ is non-linear but monotone.\\ 
Further, to investigate the effect of the dimension $p$, we consider a scenario where $p$ has relatively small value of $5$ and where it has a larger value of $10$. For comparison purposes, we choose two alternative tests: Test 1 and Test 2 are score and likelihood based tests from \cite{kong2016classical}. These alternative tests assume the following model $$ Y_i=\alpha+\int X_i(t)\beta(t)dt+\epsilon_i,  $$ where, $\epsilon_i$ are independently and identically distributed normal variables with mean $0$ and variance $\sigma^2.$ The Karhunen-Lo{\'e}ve expansion is used to represent $X_i(t)=\sum_{i=1}^{\infty} \xi_{ij}\phi_j(t),$ where the functional principal scores are $ \xi_{ij}=\int \{X_i(t)-E(X_i(t))\}\phi_j(t)dt.$ Using this expansion, the model can re-written as $Y_i=\sum_{j=1}^{\infty} \xi_{ij}\beta_j+\epsilon_i,$ where $\beta_j=\int\phi_j(t)\beta(t)dt.$ The problem of infinite parameters is addressed by the approximating the infinite dimensional model with a series of truncated models as $Y_i=\sum_{i=1}^{s_n} \xi_{ij}\phi_j(t)+\epsilon_i. $ Thus, Test 1 and Test 2 are an extension of the classical multivariate score test and likelihood test to now accommodate a diverging number of parameters ($s_n$). We call the proposed tests as Functional U (FU) test as it's based on U-statistic. As the data is dense, we use test statistic \eqref{test}. Power of all the tests for Simulation I are reported in Tables 1--3. All the results reported in this section are based on 300 replicates and level $0.05.$

%\begin{figure}
%\begin{center}
%\includegraphics[width=3in]{fig1.pdf}
%\end{center}
%\caption{Consistency comparison in fitting surrogate model in the tidal
%power example. \label{fig:first}}
%\end{figure}
%

\begin{table}[h]
	\centering
	\caption{Simulation I- Power (in \%) for Case 1.}{%
		\begin{tabular}{|cc|ccc|ccc|} \hline %***5truept
			\multicolumn{1}{|c}{}&  \multicolumn{4}{c}{$p=5$ }  &\multicolumn{3}{c|}{$p=10$ } 	\\  \hline
			n    & $\delta$  & FU  & Test 1 & Test 2 &     FU &  Test 1 & Test 2\\ \hline
			300	&	0.00	&	03.5	&	04.6	&	04.6		&		03.6	&	05.2	&	05.6	\\
			&	0.08	&	19.6	&	35.8	&	38.1		&	19.2	&	18.0	&	18.8	\\
			&	0.10	&	38.1	&	55.0	&	55.4		&	33.6	&	16.0	&	18.4	\\ \hline	
			500	&	0.00	&	03.8&	04.6	&	05.4		&	04.4		&	06.8	&	07.2	\\
			&	0.08	&	37.3	&	61.2	&	61.2		&	31.6		&	16.4	&	16.8	\\
			&	0.10	&	65.8	&	83.8	&	83.8		&	53.6		&	26.8	&	27.2	\\ \hline	
			800	&	0.00	&	03.5	&	05.8	&	05.8		&	04.4	&	06.8	&	06.8	\\
			&	0.08	&	67.3	&	83.8	&	84.2	&		55.6		&	19.6	&	19.6	\\
			&	0.10	&	88.5	&	97.3	&	97.3	&		90.4		&	27.2	&	28.8 \\ \hline	
	\end{tabular}}\label{Table 1}
\end{table}

\begin{table}[h]
	\centering
	\caption{Simulation I- Power (in \%) for Case 2.}{%
		\begin{tabular}{|cc|ccc|ccc|} \hline %***5truept
		\multicolumn{1}{|c}{}&  \multicolumn{4}{c}{$p=5$ }  &\multicolumn{3}{c|}{$p=10$ } 	\\  \hline 
			n    & $\delta$  & FU & Test 1 & Test 2 &    FU  & Test 1 & Test 2\\  \hline
			300	&	0.00	&	02.7	&	03.8	&	04.6		&		04.0	&07.0	&07.3\\
			&	0.05	&	40.8	&	11.2	&	11.5	&		29.2	&	8.00	&	8.40	\\
			&	0.08	&	83.1	&	23.5	&	24.2	&		66.8	&	13.2	&	14.0	\\ \hline	
			500	&	0.00	&	03.5&	04.6	&	05.0	&		03.2	&	04.4	&	04.8	\\
			&	0.05	&	74.2	&	12.3	&	12.3	&		55.4	&	5.60	&	6.00	\\
			&	0.08	&	97.3	&	33.1	&	33.1	&		93.6	&	10.8	&	10.8	\\ \hline	
			800	&	0.00	&	04.2&	06.5	&	06.5	&		04.9	&	03.6	&	03.6	\\
			&	0.05	&	95.0	&	20.4	&	20.4	&		82.6	&	10.5	&	10.9	\\
			&	0.08	&	100.0	&	41.2	&	41.5	&		99.6	&	15.0	&	15.0	\\ \hline					
	\end{tabular}}\label{Table 2}
\end{table}

\begin{table}[h]
	\centering
	\caption{Simulation I: Power (in \%) for Case 3.}{%
		\begin{tabular}{|cc|ccc|ccc|} \hline %***5truept
		\multicolumn{1}{|c}{}&  \multicolumn{4}{c}{$p=5$ }  &\multicolumn{3}{c|}{$p=10$ } 	\\  \hline 
		n    & $\delta$  & FU & Test 1 & Test 2 &    FU  & Test 1 & Test 2\\  \hline
			300 & 0.00 &06.0&05.0&06.0 &    04.0	&	01.0	&	01.0\\
			&0.10  &15.0&08.0&08.0&    11.0	&	09.0	&	09.0			       \\
			&0.15  &26.0&18.0&18.0&     11.0	&	11.0	&	11.0
			\\ \hline
			500 &0.00 &03.0   &05.0&05.0&    05.0	&	03.0	&	03.0
			\\
			&0.10 &14.0&06.0&06.0	&     09.0	&	08.0	&	08.0
			\\	   
			&0.15 &26.0	&13.0&13.0&     30.0	&	14.0	&	15.0
			\\ \hline
			800 &0.00 &06.0&04.0&04.0&     04.0	&	05.0	&	05.0
			\\
			&0.10 &0.22&0.12&0.12&    25.0	&	16.0	&	16.0
			\\
			&0.15 &50.0	&33.0&	33.0&     45.0	&	32.0	&	32.0
			\\ \hline
	\end{tabular}}\label{Table 3}
\end{table}

From Tables 1--3 we can see that the Type I error is controlled at level $0.05$. For all the tests the power increases with $\delta$ and with the sample size $n$, as expected. In Case 1, the data follows the typical regression model requirements of linear relation and error distribution. Note that the alternative methods limit the dimension with respect to sample size. Considering this, it comes as no surprise that in Table \ref{Table 1} for $p=5$ the linear regression based alternatives have higher power than our method but for $p=10$, our method has higher power than the alternatives. A big advantage of our method is that it is better at detecting even linear relationships when $p$ is large enough. For Case 2, our method outperforms the alternatives by a large margin even though a linear relationship is present between the scalar and the functional variables. Thus, functional regression models are sensitive to violations regarding the error structure. In Case 3, as expected our method has higher power as the relation between the two variables is non-linear and monotone.   

In Simulation II, the functional data is sparse and irregular. A sample of $n$ functions is generated using basis functions $\{ \rho_j,\,j\geq 1\}$ and model $X(t)=\sum_{k=1}^{5}\varepsilon_{k} \rho_{k}(t)$, where $ \varepsilon_j$'s are independent with a normal distribution on an equally spaced grid  $\{c_1,...,c_{56}\}$ on the interval $[0,1]$. Each curve is sampled at $5$ random points and the location of observed values of the functional data was chosen randomly from $\{c_1,...,c_{56}\}$ to generate $G_{ij}=X_i(t_{ij})+N(0,0.2),\,j=1,...,5.$ Again, we consider three cases for $Y
$.\\
\underline{Case 1}:  Fourier basis is used to generate $X$, $\varepsilon_j \sim N(0,1) $ and  $Y = \delta\int X(t)\beta(t)dt+ N(0,1).$\\
\underline{Case 2}: Fourier basis is used to generate $X$,  $\varepsilon_j \sim N(0,1) $ and $Y=	\delta\int X(t)\beta(t)dt+exp(2).$ \\
\underline{Case 3}:  For this case, we use a monomial basis and  $ \varepsilon_j \sim N(0,0.1)$. $Y=\int (0.001)^{X(t)}dt + N(0,0.1).$ \\
Note that functional data satisfies the Gaussian assumptions in Section 2.2. We again use Test 1 and Test 2  as alternatives for comparison. Extensions to these tests for the case when functional data is sparse and irregular is also available in \cite{kong2016classical}. For our proposed test, we use the test statistic \eqref{test2} as the data is sparse. Results are shown in Table \ref{Table 4}.

\begin{table}[h]
	\centering
	\caption{Power for Simulation II (values are in \%).}{%
		\begin{tabular}{|cc|ccc|ccc|ccc|} \hline %***5truept
			\multicolumn{1}{|c}{}& \multicolumn{1}{c|}{}& \multicolumn{3}{c|}{Case 1 } &
			\multicolumn{3}{c|}{Case 2 }&  \multicolumn{3}{c|}{Case 3 }	\\  \hline 
			n & $\delta$  & FU & Test 1 & Test 2  &   FU & Test 1 & Test 2  &FU & Test 1 & Test 2  \\ \hline
			300 &0.00&04.2&03.8&03.8 & 04.2&04.6&04.6 & 04.0	&	04.0	&	04.0	\\
			&0.10 &34.7&90.3&90.7 & 75.6&50.2&51.0&  30.0	&	9.0	&	10.0
			\\
			&0.15&80.6&100 &100& 96.5&86.8 &86.9&  37.0	&	23.0	&	24.0			   \\ \hline
			500 &0.00&05.0&07.3&07.3& 01.6&05.4&05.8  &01.0&06.0&07.0			\\
			&0.10&73.4&99.6&99.6& 98.5&84.2&84.2& 30.0&22.0&	24.0   \\
			&0.15&98.8&100&100 &99.6&98.1&98.1& 59.0	&41.0&	43.0
			\\ \hline
			800 &0.00&05.0&06.9 &06.9& 02.7&03.5  &03.5&  05.0&06.0&06.0
			\\
			&0.10&96.1&100&	100 & 100&96.6&96.5& 52.0	&	36.0	&	36.0
			\\	    
			&0.15&100	&100&	100 & 100&100&100& 85.0	&	60.0	&	61.0 			\\ \hline				
	\end{tabular}}\label{Table 4}
\end{table}

Table \ref{Table 4} shows that the level is well controlled for all the tests. Furthermore, we can see that the power of both the tests increases as a function of the sample size and the effect size $(\delta)$. Again for Case 1, the alternatives perform better than our method. Our test has higher power when dimension $p$ is increased. For Case 2 and Case 3 our method outperforms the others. 

In Simulation III, we test the performance when the Gaussian assumptions are violated. A sample of  $n$ functions are generated using basis functions $\{ \rho_j,\,j\geq 1\}$ and model $X(t)=\sum_{k=1}^{5}\varepsilon_{k} \rho_{k}(t)$, where $ \varepsilon_j$'s are independent and have exponential distribution with parameter $2$. The rest of the frame-work is exactly the same as in Simulation II. The result of this simulation given in Table \ref{Table 5} show results similar to Simulation II, i.e., Type-I error is controlled, alternatives outperform our method for Case 1, but under perform compared to our method for Cases 2 and 3. The test is not very sensitive to the violation of Gaussian assumptions.

\begin{table}[h]
	\centering
	\caption{Power for Simulation III (values are in \%)}{%
	\begin{tabular}{|cc|ccc|ccc|ccc|} \hline %***5truept
		\multicolumn{1}{|c}{}& \multicolumn{1}{c|}{}& \multicolumn{3}{c|}{Case 1 } &
		\multicolumn{3}{c|}{Case 2 }&  \multicolumn{3}{c|}{Case 3 }	\\  \hline 
		n & $\delta$  & FU & Test 1 & Test 2  &   FU & Test 1 & Test 2  &FU & Test 1 & Test 2  \\ \hline
			300&0.00 &01.0&	03.5&04.5 & 06.5&05.5&05.5& 03.0	&	04.0	&	04.0
			\\ 
			% &0.08&4.5	&	12.5	&	13.0	&	&	17.0	&	12.5	&	14.0	&	&? &? &?\\ 
			&0.10&11.5	&	28.0	&	28.5		&	29.0	&	14.0	&	19.5	&		100	&	73.0	&	75.0
			\\
			&0.15&15.5	&	56.0	&	56.5		&	47.0	&	29.0	&	29.0	&		100	&	84.0	&	85.0
			\\  \hline
			500&0.00&03.5	&	06.0	&	06.0		&	03.5	&	03.5	&	03.5		&	05.0	&	07.0	&	07.0
			\\
			%  &0.08&9.0	&	29.0	&	29.0	&	&	35.0	&	12.5	&	13.5	&	&	?	&	?	&	?	\\	
			&0.10&14.0	&	46.5	&	47.5		&	46.0	&	24.0	&	24.5		&100	&	78.0	&	78.0
			\\
			&0.15& 28.5	&	87.5	&	88.5		&	80.5	&	51.5	&	54.0		&	100	&	80.0	&	80.0
			\\  \hline
			800&0.00&03.0	&	05.5	&	05.5		&	04.0	&	05.5	&	05.5		&04.0	&	05.0	&	05.0
			\\
			% &0.08&13.5	&	47.0	&	47.0	&	&	67.0	&	26.0	&	26.0	&	&	?	&	?	&	?	\\
			& 0.10&26.0	&	72.5	&	73.5		&	79.5	&	36.0	&	36.0		&	100	&	75.0	&	75.0
			\\	
			&0.15& 56.5	&	97.0	&	97.0		&	98.5	&	82.0	&	82.5		&	100	&	79.0	&	79.0
			\\	 \hline	   	           		     	      		    	   
	\end{tabular}}\label{Table 5}
\end{table}

The advantages of the proposed statistic are clearly demonstrated through these simulations. The alternatives perform well when their model assumptions regarding the linearity of relation between the two variables, controlled number of parameters and the normality of error are satisfied. Violation of any of these leads to a significant deterioration of power. In this regard, the FU test is relatively robust - a highly desirable property as data applications seldom guarantee that these assumptions are satisfied.

\begin{center}
	{\section{DATA APPLICATION}}
\end{center}	
We provide analysis of two data sets - Diffusion Tensor Imaging data and Ebay data to test our method on the dense and sparse data respectively. 
\subsection{Diffusion Tensor Imaging}
White matter is a tissue in the brain composed of nerve fibers. These fibers (axons) are protected by myelin, a type of fat. Multiple sclerosis (MS) is a central nervous system disorder in which the white blood cells enter the nervous system injuring it. The myelin sheath that protects the axons which are responsible for propagating electrical signals in the brain is stripped off during inflammation. Thus, any damage to the sheath can can cause cognitive and motor disability. In this analysis, we want to test whether differences in cognitive functions are associated with changes in white matter of patients with MS. In particular, we focus on the intracranial white matter in the corpus callosum (CCA) and the
right corticospinal tract (RCST). Diffusion tensor imaging (DTI) tractography is a magnetic resonance imaging technique that allows the study of white-matter tracts by measuring the diffusivity of water in the brain. From these images,
continuous summaries of major white matter structures called
tract profiles can be obtained, which are functional variables. Cognitive function is quantified via the Paced Auditory Serial Addition Test (PASAT) score, which is a scalar variable. The dataset DTI is available R package REFUND and contains information in the on the CCA and RCST tract profiles  of 100 individuals with MS, scanned once approximately every year. The patients have 2-8 visits, and to obtain independence, we focus on data collected on the first visits. Further details on this data set can been found in  \cite{swihart2014restricted} and \cite{goldsmith2011penalized}.  We consider the following tests: $H_{01}$: there is no association between the PASAT score and the white matter in CCA tract; and $H_{02}$: there is no association between the PASAT score and the white matter in RCST tract. The proposed approach generates test statistic $37.17$ and p-value$<0.001$ for $H_{01}$ and test statistic $26.46 $ and p-value=0.35 for $H_{02}$. Thus, there is evidence for association of cognitive ability with white matter tract profiles in CCA but no evidence of its association with white matter tract profiles in RCST. The conclusions of the alternative tests were also the same. There is evidence in the literature that this association is present in MS patients \citep{ozturk2010mri} confirming the validity of our findings.

%Processes of the central nervous system such as aging influence the microstructural composition and architecture of the affected tissues. This in turn affects the diffusion of water within the tissues which can be captured/measured by diffusion tensor imaging (DTI). In this analysis, the association between white matter tracts and cognitive function in patients with multiple sclerosis (MS) is of interest. The analyzed DTI dataset is available in the R package REFUND. The data consists of the DTIs of white matter in the corpus callosum (CCA) and right cortico spinal tract (RCST) of 100 individuals with MS, scanned once approximately every year. The patients have 2-8 visits, and to obtain independence, we focus on data collected on the first visits. The diffusivity measures include the mean diffusivity for the CCA and parallel diffusivity for the RCST. In addition, the Paced Auditory Serial Addition Test (PASAT) score, which is a scalar measure of cognitive function, is also present. We consider the following tests: $H_{01}$: there is no association between the PASAT score and CCA tract; and $H_{02}$: there is no association between the PASAT score and RCST tract. The proposed approach generates test statistic $37.17$ and p-value$<0.001$ for $H_{01}$ and test statistic $26.46 $ and p-value=0.35 for $H_{02}$. The finding is consistent with Swihart et al. (2014), which takes different analysis strategies.

\subsection{Ebay Auction Data}
Online auctions are very popular due to their flexibility and convenience. It is of obvious economic interest to study and understand the dynamics of auctions. One such aspect of interest is the determination of the different factors that influence the price of an item during the course of the auction \citep{WolfgangJank}. Using the data obtained from the 7-day auction of Palm M515 PDA (personal digital assistants) on eBay during the period of March 2003 - May 2003, we investigate whether the opening price of Palm M515 PDA impacts its price trajectory. This data is  freely available online (http://www.modelingonlineauctions.com/datasets). It contains information on WTP (willing-to-pay) or proxy bids instead of real-price or live bids placed by a bidder. The proxy bids are first converted to the real price, the details of which can be found in \cite{liu2008functional}. After removing auctions with two or fewer bids, the remaining data contained information on 145 Palm M515 devices. The price histories of these devices can be viewed as i.i.d realizations of an underlying price process \citep{liu2008functional}. As the bids typically arrive slowly at the start of an auction and rapidly towards the end, the price process is observed on a sparse and irregular grid. Hence, we applied the test based on test statistic \eqref{test2} , to reject the null hypothesis of no association ($p$-value$<10^{-4}$). The $p$-value for Test 1 was $0.065$ and that for Test 2 was $0.052.$ Thus, the alternatives found no association between the opening price and price trajectory. Online auction literature suggests that the opening price does have an effect on the price trajectory particularly at the start of the trajectory \citep{WolfgangJank}. Thus, our method was able to detect the association while the alternatives could not.

\begin{center}
	{\section{Conclusion}}
\end{center}

In this article, we have provided a robust method to determine association between a scalar and a functional variable. Most existing alternatives are based on linear models and suffer a loss in power when the underlying assumptions such as non-linear relation between the variables are violated. We also extend our method to the case when functional data is sparse and show that it performs well even when the Gaussian assumptions are disobeyed. Given that we obtain approximate asymptotic distribution the proposed method is computationally easy and quick to implement. It can serve as a useful tool for data exploration in data analysis. It is prudent to check whether association exists between variables before modeling their relation. It may be used as a tool for preliminary variable selection in functional models. Thus, our method offers great practical utility.

Some caution does need to be taken during the implementation of this method. We have implicitly assumed that the functional data is aligned which is the norm in Functional Data literature. If the data is not aligned, then to avoid loss of power the data needs to be aligned before testing. Several curve registration techniques are available in the literature. It does not accommodate covariates. It is worthwhile to investigate extensions of this method to account for covariates. Other possible extensions are to test associations between two functional variables, or between a vector of scalar variables and a vector of functional variables. Thus, there is a wide scope for extensions of the proposed method and consequently increasing its applicability. 
%Our method is able to detect more general relations namely monotone relations and thus, has a greater practical utility.   

\begin{center}
	{\section*{Acknowledgments}}
\end{center}
“The MRI/DTI data were collected at Johns Hopkins University and the Kennedy-Krieger Institute"

%\bigskip
%\begin{center}
%{\large\bf SUPPLEMENTARY MATERIAL}
%\end{center}

%\section{BibTeX}

\begin{center}{\section*{Appendix }}
\end{center}
We first obtain some results necessary for Theorem 2. All the limits are taken as $n\to \infty$.

\begin{proposition}\label{prop2}
	We have $\underset{1 \leq i \leq n }{sup} \|V_i-W_i\| \to 0,$ almost everywhere (a.e).	
\end{proposition}	 

%\begin{proof}
\textbf{Proof } \\For a fixed $t \in I$,  	
	\begin{eqnarray*}
		|V_i(t)-W_i(t)| 
		&\leq | E[ \mathcal{I}(y_i>Y,x_i(t)>X(t))]-n^{-1}\sum_{j=1}^{n} \mathcal{I}[y_i>Y_j,x_i(t)>X_j(t)]|\\
		& \quad +|E[ \mathcal{I}(y_i<Y)\mathcal{I}(x_i(t)<X(t))]-n^{-1}\sum_{j=1}^{n} \mathcal{I}(y_i<Y_j)\mathcal{I}(x_i(t)<X_j(t))|
	\end{eqnarray*}
	
	Using the Glivenko-Cantelli	for multivariate variables \citep{shorack2009empirical}, 
	$ \underset{1 \leq i \leq n }{sup}|V_i(t)-W_i(t)| \to 0 ,$ almost everywhere,
	which yields, $ \underset{1 \leq i \leq n }{sup}|V_i(t)-W_i(t)|^2 \to 0 , $ a.e.
	Note that, both $V_i$ and $W_i$ are uniformly bounded. Thus, using the Lebesgue dominated convergence theorem $$\underset{n \to \infty }{lim} \int   \underset{1 \leq i \leq n }{sup}|V_i(t)-W_i(t)|^2dt =\int\underset{n \to \infty }{lim} \underset{1 \leq i \leq n }{sup}|V_i(t)-W_i(t)|^2dt=0,\, a.e. $$ This implies,
	$\underset{1 \leq i \leq n }{sup} \|V_i-W_i\|= \underset{1 \leq i \leq n }{sup} \int |V_i(t)-W_i(t)|^2 dt \leq \int  \underset{1 \leq i \leq n }{sup}|V_i(t)-W_i(t)|^2dt \to 0,\,a.e.,  $ proving the result.
	
%\end{proof}

\begin{proposition}
	$ E\| \mathcal{C}_v- \mathcal{C}_w\|_S^2 \to 0  $
\end{proposition}

%\begin{proof}
\textbf{Proof } \\
	Consider,
	\begin{align*}
	\| \mathcal{C}_v- \mathcal{C}_w\|_S^2 &=\left(\sum_{k=1}^{\infty} \|\mathcal{C}_v(e_k)- \mathcal{C}_w (e_k) \|^2 \right) \\
	&=\left(\sum_{k=1}^{\infty}\left\|\sum_{i=1}^{n}\dfrac{\langle V_i,e_k\rangle V_i}{n}-\dfrac{\langle W_i,e_k\rangle W_i}{n}    \right\|^2 \right)\\
	%	&=E\left(\sum_{k=1}^{\infty}\langle \sum_{i=1}^{n}\dfrac{ \langle V_i,e_k  \rangle V_i}{n}- \dfrac{\langle W_i,e_k\rangle W_i}{n},  \sum_{j=1}^{n}\dfrac{ \langle V_j,e_k  \rangle V_j}{n}- \dfrac{\langle W_j,e_k\rangle W_j}{n}    \rangle     \right)\\
	%	&=En^{-2}\left(\sum_{k=1}^{\infty}  \sum_{i,j=1}^{n}  \langle \langle V_i,e_k  \rangle V_i-\langle W_i,e_k\rangle W_i,  \langle V_j,e_k  \rangle V_j-\langle W_j,e_k\rangle W_j   \rangle   \right)\\
	%	&=En^{-2}\left(\sum_{k=1}^{\infty}  \sum_{i,j=1}^{n}  \langle V_i,e_k  \rangle \langle V_j,e_k  \rangle \langle V_i,V_j\rangle- \langle V_i,e_k  \rangle \langle W_j,e_k  \rangle \langle V_i,W_j\rangle   \right) \\
	%	&\quad + En^{-2}\left(\sum_{k=1}^{\infty}  \sum_{i,j=1}^{n}  \langle W_i,e_k  \rangle \langle W_j,e_k  \rangle \langle W_i,W_j\rangle- \langle W_i,e_k  \rangle \langle V_j,e_k  \rangle \langle W_i,V_j\rangle   \right) \\
	&=n^{-2}\left(\sum_{k=1}^{\infty}  \sum_{i,j=1}^{n}  \langle V_i,e_k  \rangle (\langle V_j,e_k  \rangle \langle V_i,V_j\rangle- \langle W_j,e_k  \rangle \langle V_i,W_j\rangle  ) \right) \\
	&\quad + n^{-2}\left(\sum_{k=1}^{\infty}  \sum_{i,j=1}^{n}  \langle W_i,e_k  \rangle (\langle W_j,e_k  \rangle \langle W_i,W_j\rangle-  \langle V_j,e_k  \rangle \langle W_i,V_j\rangle)   \right)\\
	&=A+B
	\end{align*}
	Consider, 
	\begin{align*}
	A&=n^{-2}\left(\sum_{k=1}^{\infty}  \sum_{i,j=1}^{n}  \langle V_i,e_k  \rangle (\langle V_j,e_k  \rangle \langle V_i,V_j\rangle- \langle W_j,e_k  \rangle \langle V_i,W_j\rangle  ) \right)\\
	%	&=En^{-2}\left(\sum_{k=1}^{\infty}  \sum_{i,j=1}^{n}  \langle V_i,e_k  \rangle (\langle V_j,e_k  \rangle \langle V_i,V_j\rangle- \langle V_j,e_k  \rangle  \langle V_i,W_j\rangle   ) \right)\\
	%	&\quad+ En^{-2}\left(\sum_{k=1}^{\infty}  \sum_{i,j=1}^{n}  \langle V_i,e_k  \rangle (\langle V_j,e_k  \rangle  \langle V_i,W_j\rangle- \langle W_j,e_k  \rangle \langle V_i,W_j\rangle  ) \right)\\
	&=n^{-2}\left(\sum_{k=1}^{\infty}  \sum_{i,j=1}^{n}  \langle V_i,e_k  \rangle \langle V_j,e_k  \rangle( \langle V_i,V_j\rangle- \langle V_i,W_j\rangle   ) \right)\\
	&\quad+ n^{-2}\left(\sum_{k=1}^{\infty}  \sum_{i,j=1}^{n}  \langle V_i,e_k  \rangle\langle V_i,W_j\rangle (\langle V_j,e_k  \rangle  - \langle W_j,e_k  \rangle  ) \right)\\
	&=A_1+A_2
	\end{align*}	
	
	Consider, 
	\begin{align*}
	A_1&=n^{-2}\left(\sum_{k=1}^{\infty}  \sum_{i,j=1}^{n}  \langle V_i,e_k  \rangle \langle V_j,e_k  \rangle( \langle V_i,V_j\rangle- \langle V_i,W_j\rangle   ) \right)\\
	&\leq n^{-2}\sum_{k=1}^{\infty}  \sum_{i,j=1}^{n} |  \langle V_i,e_k  \rangle |   | \langle V_j,e_k  \rangle   | | \langle V_i,V_j-W_j\rangle|\\
	%	& \leq  En^{-2} \sum_{i,j=1}^{n}\| V_j-W_j\| \left(\sum_{k=1}^{\infty} \langle V_i,e_k  \rangle | ^2  \sum_{k=1}^{\infty}| \langle V_j,e_k  \rangle   |^2\right)^{(1/2)}\\
	&\leq n^{-2} \sum_{i,j=1}^{n}\| V_j-W_j\| \|V_j\| \|V_i\| \\
	& \leq c (\underset{j }{sup}\| V_j-W_j\| )\\
	%	& \to 0
	\end{align*}
	Taking expectation and using the fact that 	$V_j$ and $W_j$ are bounded uniformly we obtain $E(A_1)\to 0$. Similarly we can show $E(A_2) \to 0$ and $E(B) \to 0$ to prove the proposition.
%\end{proof}	

\textbf{Proof of Theorem \ref{thm2}}:

%\begin{proof}
	Theorem 2.5 from \cite{horvath2012inference} states that $E\| \mathcal{C}_v -\mathcal{C} \|^2_s \to 0$. This along with Proposition 1 proves Theorem 2.\\
%\end{proof}	

\textbf{Proof of Theorem \ref{thm3}}:
%\textbf{Proof } \\
%\begin{proof}
This proof uses fact that, for all $t \in I$, $\widehat{X}_i^K(t) \to \widetilde{X}_i(t),\, n \to\infty,\, K \to \infty $ in probability. All the limits in the following are taken as $n \to \infty$ and $K \to \infty$. By Urysohn's lemma, $\mathcal{I}(\widehat{X}_i(t)>0) $ can be approximated by a continuous function $f(\widehat{X}_i(t))$. Using this, it can be easily showed that $\mathcal{I}(\widehat{X}_i(t)>0) \to \mathcal{I}(\widetilde{X}_i(t)>0) $ in probability for all $t \in I $.  Let $\widetilde{U}_{ij}(t)=\mathcal{I}[(Y_i-Y_j)(\widetilde{X}_i(t)-\widetilde{X}_j(t))>0] $ and $\widehat{U}^K_{ij}(t)=\mathcal{I}[(Y_i-Y_j)(\widehat{X}^K_i(t)-\widehat{X}^K_j(t))>0].$ Combining the above with the fact that $\mathcal{I}[(Y_i-Y_j)(\widetilde{X}_i(t)-\widetilde{X}_j(t))>0]=\mathcal{I}(Y_i-Y_j>0)\mathcal{I}[\widetilde{X}_i(t)-\widetilde{X}_j(t)>0]+\mathcal{I}(Y_i-Y_j<0)\mathcal{I}[\widetilde{X}_i(t)-\widetilde{X}_j(t)<0],$ we obtain $ \widehat{U}^K_{ij}(t)  \to \widetilde{U}_{ij}(t)$ in probability for all $t$. As $\widehat{U}^K_{ij}(t)$ are bounded for all $i,j$ we have $ E[\widehat{U}^K_{ij}(t)- \widetilde{U}_{ij}(t)]^2 \to 0 $ which further yields $\underset{ij}{sup} \mid \widehat{U}^K_{ij}(t)- \widetilde{U}_{ij}(t)\mid \to 0$ in probability. Thus, $\widehat{U}_n^K(t) \to \widetilde{U}(t)$ in probability for all $t$ which mean $\widehat{T}^K \to \widetilde{T} $ in probability. 
	
%\end{proof}	

\end{document}